\documentclass[conference]{IEEEtran}
 
\usepackage{cite}

\ifCLASSINFOpdf
  \usepackage[pdftex]{graphicx}
\else
\fi
\usepackage{amsmath}
\usepackage{url}

\usepackage[]{fancyhdr} %
\newcommand{\changefont}{\fontsize{9}{9}\selectfont}
\ifCLASSOPTIONcompsoc
  \usepackage[caption=false,font=normalsize,labelfont=sf,textfont=sf]{subfig}
\else
  \usepackage[caption=false,font=footnotesize]{subfig}
\usepackage{circuitikz}
\usepackage{amssymb}

\IEEEoverridecommandlockouts
\begin{document}

%
\title{Intricacies of formulating measurement-based small signal models of inverter based resources}

\author{\IEEEauthorblockN{Sushrut Thakar}
\IEEEauthorblockA{School of Electrical, Computer and Energy Engineering\\Arizona State University\\
Tempe, USA\\
sushrut.thakar@asu.edu}
\and
\IEEEauthorblockN{Deepak Ramasubramanian}
\IEEEauthorblockA{Power Delivery and Utilization Department\\Electric Power Research Institute\\
Knoxville, USA\\
dramasubramanian@epri.com} \thanks{This work was funded by EPRI Research Program P173A: Modeling and Analytics with Emerging Technologies.}}


%





\maketitle
\thispagestyle{fancy}
\pagestyle{fancy}


\begin{abstract}
The increasing number of inverter based resources (IBRs) connected to modern power systems necessitate accurate modeling of IBRs in stability studies. Measurement-based modeling approaches can create an impedance/admittance model of IBRs to be subsequently used in small signal stability studies. Such measurement-based approaches have previously been successfully used for dc-dc converters but present unique challenges for three-phase IBRs. This paper discusses the choices and challenges that can lead to an incorrect model from measurement-based approaches as well as mitigation mechanisms of the adverse impacts using an illustrative example. A discussion of small signal stability assessment involving such admittance models is also provided. It is seen through the illustrative example shown that the injection signal and the model fitting procedure, the phase locked loop used for the measurement device, and the grid impedance may impact the accuracy of the measurement-based admittance model of the IBRs.
\end{abstract}



%
\IEEEpeerreviewmaketitle

\section{Introduction}\label{sec:intro}
In recent years, there has been an increasing number of inverter based resources (IBRs) being installed in the power grids at various levels. Several cases of oscillations or network disturbances have been shown to have either resulted or exacerbated by the increased presence of IBRs in the grid~\cite{philinv}. Consequently, researchers have developed detailed small signal models for IBRs~\cite{ghe13,ghe14} in order to incorporate IBRs into stability studies. These models rely on detailed modeling of all the control blocks and components which comprise the IBR. However, many of the IBRs installed in the field rely on off-the-shelf inverter systems - and the detailed knowledge about the exact control structure and parameters implemented may not be known. Hence, a different approach is often needed to estimate/build the inverter models. Such an approach comes in the form of measurement-based models, which can be created either using the measurements carried out on the physical device (inverter) itself or using a black-box model provided by the inverter manufacturer. \\

Such measurement-based approaches to estimate the inverter output admittance have been successfully used for dc-dc converters in the past~\cite{init6}. However, applying similar approaches for three-phase IBRs provides unique challenges. Since an inverter operates on alternating current, the currents and voltages do not have a constant steady state value but oscillate with the grid frequency. Further, there are three terminal voltages and currents for a three-phase IBR. Hence, IBR models in a $dq$ domain rotating with the same grid frequency~\cite{ghe13} are most prominently used and are considered in the present paper. IBR models in the phase and sequence domains are interconvertible with the models in $dq$ domain~\cite{impedanceeqv}.\\

While creating a measurement-based impedance/admittance model of an IBR, there are several pitfalls and considerations which can lead to an incorrect impedance model. These include choosing the variable for the small signal injection~\cite{ghe23}, the choice of the injection signal~\cite{hgthesis,ghe21,ghe41,ghe49}, choosing the magnitude and the duration of the injection signals~\cite{ghe28,ghe29,ghe30,ghe41,ghe49}, removing the influences of the phase locked loop (PLL)~\cite{ghe38,ghe41,gfthesis} and the grid impedance~\cite{ghe38,ghe41}, and model fitting~\cite{ghe42,ghe51}. Many of the references available chiefly concentrate on one particular aspect mentioned, whereas the present paper aims to give an overview of various choices and considerations which must be made while creating a measurement-based model. These are discussed with the help of an illustrative inverter model/example. The paper highlights how various impacts/choices might lead to the creation of an incorrect inverter model, including choices such as the magnitude of the injection signal, which are not discussed in detail in the literature. This paper also compares various approaches available in the literature to mitigate the impact of the measurement PLL and grid impedance.\\

The rest of this paper is organized as follows: Section \ref{sec:invmodel} describes the inverter model used to develop the measurement-based procedure and for the validation. Section \ref{sec:measdesign} describes the aspects related to applying the disturbance to the models and measurement of the model response. Section \ref{sec:considerations} describes various choices and considerations involved while creating the measurement-based models, while Section \ref{sec:pllgrid} describes mitigating the impacts of the measurement phase locked loop and the grid. Section \ref{sec:stability} studies some stability characteristics of the constructed model, and Section \ref{sec:conclusion} concludes the paper.

\section{Inverter model}\label{sec:invmodel}
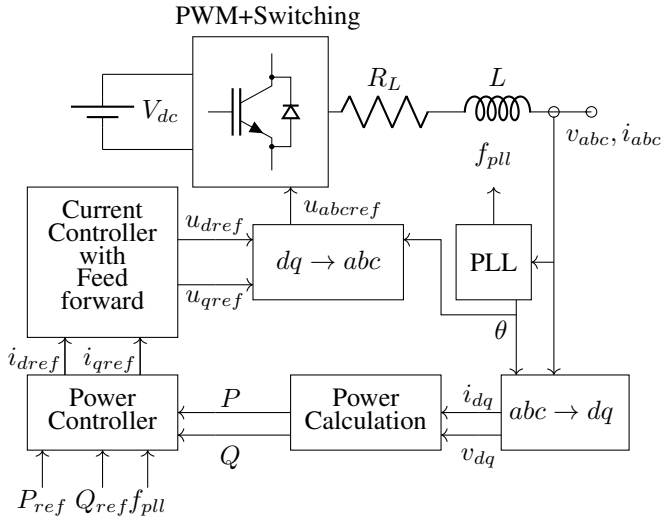
\begin{figure}[!t]
	\centering
	\begin{circuitikz}[american voltages]
		\draw (-1,1.5) to [battery1, l=$V_{dc}$] (-1,0.5);
		\draw (-1,1.5) to (0.2,1.5);
		\draw (-1,0.5) to (0.2,0.5);
		\draw (0.2,0) rectangle (2,2);
		\draw (1.25,1) node[nigbt,bodydiode]{};
		\draw (1.2,2.25) node{PWM+Switching};
		\draw (2,1) to[R=$R_L$] (3.5,1)
		to[L=$L$] (5,1);
		\draw (5,1) to (5.5,1) circle(2pt);
		\draw (5,1) circle(2pt);
		\draw[->] (5,0.94)to[->] (5,-2.5);
		\draw (5.8,0.7) node []{$v_{abc},i_{abc}$};
		\draw[->] (5,-1) to (4.7,-1);
		\draw (4.7,-0.5) rectangle (3.7,-1.5);
		\draw (4.2,-1) node[]{PLL};
		\draw[->] (4.5,-1.5) to (4.5,-2.5);
		\draw (4.3,-1.9) node[]{$\theta$};
		\draw[->] (4.5,-1.7) to (3.5,-1.7) to (3.5,-0.7) to (3,-0.7);
		\draw[->] (4.2,-0.5) to ++(0,0.5) node[label=above:{$f_{pll}$}]{};
		\draw (4.3,-2.5) rectangle (6,-3.5);
		\draw (5.1,-3) node[]{$abc\rightarrow dq$};
		\draw (4.3,-3) to (4,-3);
		\draw (4.0,-2.8) node[]{$i_{dq}$};
		\draw (4.3,-3.3) to (4,-3.3);
		\draw (4.0,-3.6) node[]{$v_{dq}$};
		\draw[->] (4,-3) to (3.5,-3);
		\draw[->] (4,-3.3) to (3.5,-3.3);
		\draw (1.5,-2.5) rectangle (3.5,-3.5);
		\draw (2.5,-2.8) node[]{Power};
		\draw (2.5,-3.1) node[]{Calculation};
		\draw (1.5,-3) to (0.5,-3);
		\draw (0.7,-2.8) node[]{$P$};
		\draw (1.5,-3.3) to (0.5,-3.3);
		\draw (0.7,-3.6) node[]{$Q$};
		\draw[->] (0.5,-3) to (0,-3);
		\draw[->] (0.5,-3.3) to (0,-3.3);
		\draw (0,-2.5) rectangle (-2,-3.5);
		\draw (-1,-2.8) node[]{Power};
		\draw (-1,-3.1) node[]{Controller};
		\draw[->] (-0.4,-4.0) to ++(-0.0,0.5);
		\draw (-0.4,-4.2) node[]{$f_{pll}$};
		\draw[->] (-1.8,-4.0) to ++(-0.0,0.5);
		\draw (-1.8,-4.2) node[]{$P_{ref}$};
		\draw[->] (-1.0,-4.0) to ++(-0.0,0.5);
		\draw (-1.0,-4.2) node[]{$Q_{ref}$};
		\draw (-0.5,-2.5) to (-0.5,-2);
		\draw (-0.9,-2.3) node[]{$i_{qref}$};
		\draw (-1.5,-2.5) to (-1.5,-2);
		\draw (-1.9,-2.3) node[]{$i_{dref}$};
		\draw[->] (-0.5,-2.5) to (-0.5,-2);
		\draw[->] (-1.5,-2.5) to (-1.5,-2);
		\draw (0,-2) rectangle (-2,0);
		\draw (-1,-0.3) node[]{Current};
		\draw (-1,-0.6) node[]{Controller};
		\draw (-1,-0.9) node[]{with};
		\draw (-1,-1.2) node[]{Feed};
		\draw (-1,-1.5) node[]{forward};
		\draw (0,-1.3) to (0.5,-1.3);
		\draw (0.5,-1.5) node[]{$u_{qref}$};
		\draw (0,-0.7) to (0.5,-0.7);
		\draw (0.5,-0.5) node[]{$u_{dref}$};
		\draw[->] (0.5,-1.3) to (1,-1.3);
		\draw[->] (0.5,-0.7) to (1,-0.7);
		\draw (1,-0.5) rectangle (3,-1.5);
		\draw (2,-1) node[]{$dq\rightarrow abc$};
		\draw[->] (1.5,-0.5) to (1.5,0);
		\draw (2.2,-0.3) node[]{$u_{abcref}$};
	\end{circuitikz}
	\caption{Block diagram of the inverter model}
	\label{fig:invblockdig}
\end{figure}
A three-phase inverter model was created in PSCAD to illustrate various aspects related to creating a measurement-based model. The control of this inverter was chosen to be modeled in the $dq$ frame. The overall block diagram of the inverter is shown in Fig. \ref{fig:invblockdig}. Here, the active power is controlled based on the active power reference set in the model and the frequency deviation from the nominal frequency of 60 Hz. The reactive power is controlled based on the reactive power reference given to the model. An average model of the inverter switching is used to represent the inverter switching action. This model hence ignores the dynamics associated with the pulse width modulation and switching. However, typically the switching takes place at a high frequency (several kHz) and the switching dynamics largely occur outside the region of frequency concentrated on in this paper (less than 200 Hz), hence this is an acceptable assumption. All the controllers used in various blocks are proportional-integral (PI) controllers. The control and electrical parameters used are listed in Table \ref{tab:paramvals}. The operating point selected has a terminal voltage of 1.01 p.u. and active and reactive power injection from the inverter of 195 MW and 40 MVAr. The grid is modeled as an ideal voltage source behind an impedance involving both resistive and reactive components. This structure can represent an infinite bus if the grid impedance is made zero (or infinitesimally small), or it can represent a strong or weak grid depending on the magnitude of the grid impedance. A larger grid impedance signifies a weaker grid. To validate the measurement based admittance model, an analytical small signal model of the inverter was formed using a state-space form by using sympy~\cite{sympy} and python-control~\cite{python-control} python packages.\\

\begin{table}[!t]
	\caption{Values of the inverter parameters}
	\label{tab:paramvals}
	\centering
	\begin{tabular}{|l | c| l |}
		\hline
		Parameter & Value & Description\\ [0.5ex]
		\hline
		$f$ & 60.0 & The nominal system frequency (Hz) \\
		$S_{base}$ & 200.0 & The system MVA base \\
		$V_{base}$ & 120.0 & The system voltage base (kV LL RMS)\\
		$L$ & 0.15 & Inductance of the coupling inductor (p.u.)\\
		$R_L$ & 0.0015 & Resistance of the coupling inductor (p.u.)\\
		$K_{p,i}$ & 0.5 & Proportional term in the current controller\\
		$K_{i,i}$ & 20 & Integral term in the current controller\\
		$K_{p,pll}$ & 20 & Proportional term in the PLL controller\\
		$K_{i,pll}$ & 700 & Integral term in the PLL controller\\
		$K_{p,p}$ & 0.5 & Proportional term in the active power controller\\
		$K_{i,p}$ & 20 & Integral term in the active power controller\\
		$K_{p,q}$ & 0.5 & Proportional term in the reactive power controller\\
		$K_{i,q}$ & 20 & Integral term in the reactive power controller\\
		$K_{drp}$ & 20 & Frequency-active power droop term\\ [1ex]
		\hline
	\end{tabular}	
\end{table}

The measurement-based model was created by considering the described inverter model at the designated operating point as a black-box model according to the procedure described in the next section. The rest of the paper then discusses the impact of various factors on measurement-based model creation. Note that the values of various parameters for which the measurement-based model is valid/correct are dependent on the inverter black-box model as well as the operating point. Hence for a different black-box model, the values of various parameters for which the measurement-based approach leads to a correct model can differ, and more work would be needed. However, the same general principles would govern, and the discussion about various choices/parameters presented would still be applicable.

\section{Measurement-Based Model Creation}\label{sec:measdesign}
The idea behind the measurement-based small signal model is to inject a number of small disturbances to the black-box model/inverter and measure and store its responses. Once enough data is collected, numerical fitting techniques can be used to build a parametric model for the admittance or impedance of the inverter, looking from the grid. Such admittance or impedance models can then be used for small signal stability analysis~\cite{ghe10}.\\

A quantity oscillating in the $abc$ frame with the grid frequency becomes a dc quantity in the $dq$ domain. Further, the number of quantities is reduced to 2 (along the $d$ and $q$ axes - ignoring the 0-axis) instead of 3 (along the $a$, $b$, and $c$ axes). With two axes, the represented admittance will form a 2-by-2 matrix, represented as a multiple-in, multiple-out (MIMO) control system.
The individual terms of this 2-by-2 transfer function are defined as follows:
\begin{align}
	Y_{inv} = 
	\begin{pmatrix}
		Y_{dd} & Y_{dq}\\
		Y_{qd} & Y_{qq}\\
	\end{pmatrix}
	=
	\begin{pmatrix}
		\frac{\partial i_d}{\partial v_d} & \frac{\partial i_d}{\partial v_q}\\
		\frac{\partial i_q}{\partial v_d} & \frac{\partial i_q}{\partial v_q}
	\end{pmatrix}\label{eq:Ydef}
\end{align}

In general, due to the presence of inductive elements and various controls that are common to the $d$ and $q$ axis, all four terms of this admittance can be expected to be non-zero. The general procedure followed while analyzing the data to estimate the admittance model is as follows:
\begin{enumerate}
	\item Save the steady-state snapshot of the model without any injection.
	\item Select the set of frequencies at which the inverter admittance is to be estimated.
	\item Inject a signal of the suitable magnitude of each selected frequency twice, once on the $d$ axis and then on the $q$ axis, and record the inverter terminal voltages and currents in each case. For all the cases, also store the output of the measurement PLL controller passed through a high pass filter (``$d\theta_{filtered}$'' ) as mentioned in Section \ref{sec:pllgrid}.
	\item Correct the measured $d$ and $q$ axis voltages and currents by using the stored $d\theta_{filtered}$ to remove the influence of the measurement PLL.
	\item For each frequency, convert the time domain voltage and current signals to the frequency domain signals using FFT. Retain the magnitude and phase data for the injected frequency component.
	\item For each frequency in the set of the injected frequencies, using the obtained Fourier components of voltages and currents, obtain the 2-by-2 admittance matrix.
	\item Once the 2-by-2 admittance matrix is estimated for all the frequencies in the set of the injected frequencies, use vector fitting to estimate the transfer functions (parametric model) of the inverter admittance.
\end{enumerate}

Out of these steps, step 1 is entirely performed in PSCAD. Step 2 is an input to be decided. Step 3 is completed using the python interface of PSCAD and steps 4-7 are completed entirely in python. For numerical calculations, scipy~\cite{Scipy} and numpy~\cite{numpy} packages are used. For step 7, `VectorFitting' module from the python package scikit-rf~\cite{skrf} is used. Here, a prerequisite for conducting the measurements is that the model needs to reach a steady state before applying the small signal disturbance. Hence, neither the inverter model itself nor the inverter model combined with the grid at the designated operating condition should be small signal unstable while conducting the measurements.\\

\begin{figure}[!t]
	\centering
	\includegraphics[width=0.5\textwidth]{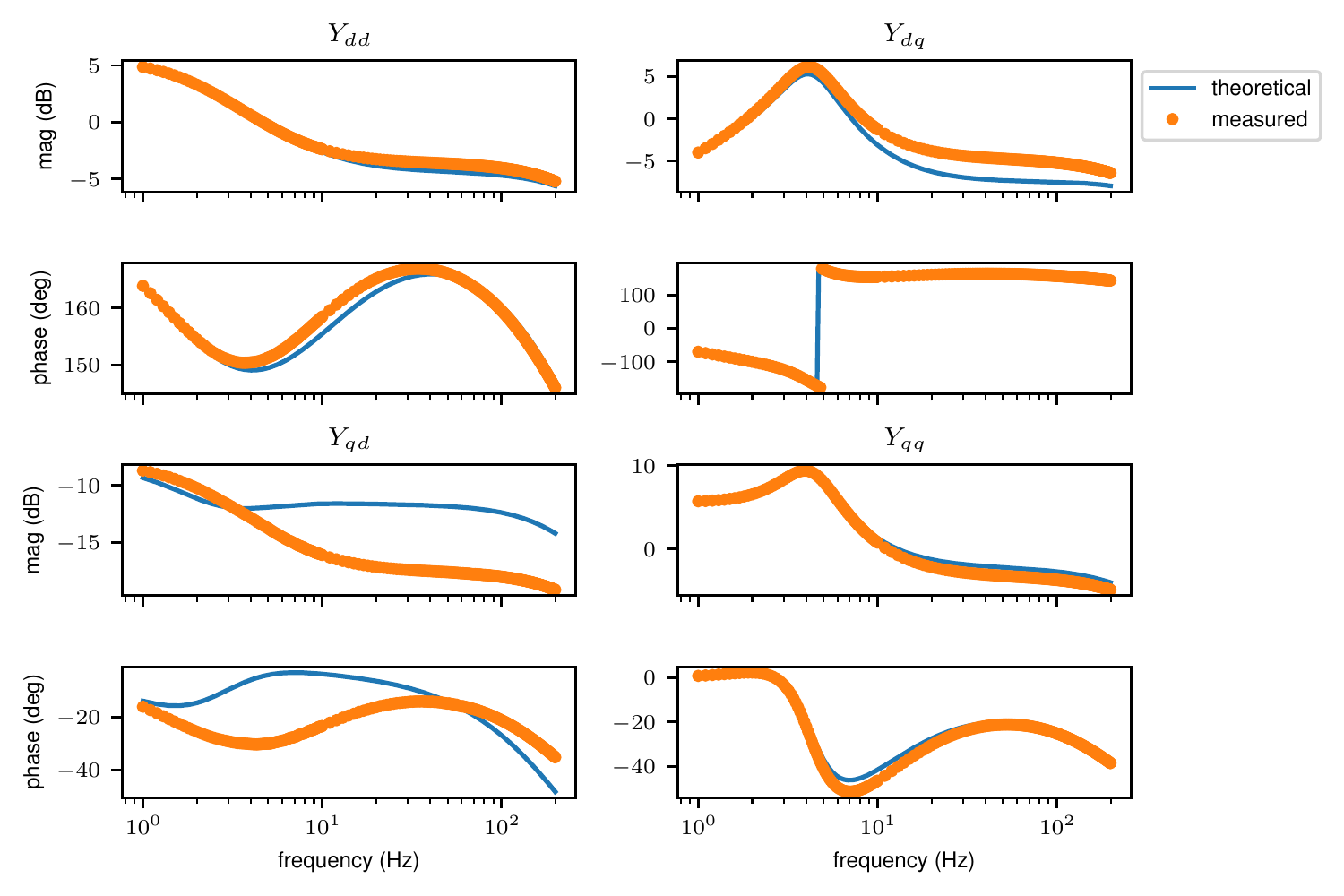}
	\caption{Small signal versus measurement-based transfer function of the inverter admittance. The magnitude and phase plots at the top-left, bottom-left, top-right and bottom-right correspond to $Y_{dd}$, $Y_{qd}$, $Y_{dq}$ and $Y_{qq}$ respectively.}
	\label{fig:Nonparametric}
\end{figure}

Fig. \ref{fig:Nonparametric} shows the comparison between the transfer function obtained from the small signal model versus the estimated transfer function by following steps 1-6 mentioned at the beginning of this section.
It is seen that the estimated $Y_{dd}$, $Y_{dq}$ and $Y_{qq}$ are very close to the values obtained from the small signal model, whereas $Y_{qd}$ obtained from both methods show slightly different behavior. However, the magnitude of $Y_{qd}$ is small, hence this may be due to a numerical reason or slight difference between the PSCAD and small signal models. Further, it is shown in Subsection \ref{subsec:stabtest} that in spite of this difference, the fitted transfer function obtained shows a similar stability behavior as the small signal model, or the stability characteristics are not highly affected by the difference in $Y_{qd}$ in this case since the value of $Y_{qd}$ is small compared to the other three admittance terms.

\section{Various considerations for the measurement-based admittance estimation}\label{sec:considerations}
\subsection{Current versus voltage injection}
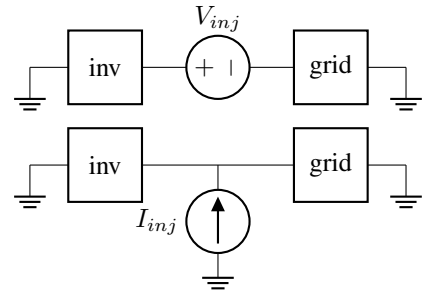
\begin{figure}[!t]
	\centering
	\begin{circuitikz}[american voltages, american currents]
		\draw (0,0) node[ground]{} to [twoport,t=inv] ++(2,0) to [vsource, l=$V_{inj}$] ++(1,0) to [twoport,t=grid] ++(2,0) node[ground]{};
		\draw (0,-1.3) node[ground]{} to [twoport,t=inv] ++(2,0) to ++(1,0) to [twoport,t=grid] ++(2,0) node[ground]{};
		\draw (2.5,-2.8) node[tlground]{} to [isource, l=$I_{inj}$] ++(0,1.5);
	\end{circuitikz}
	\caption{Voltage and current injection schematics}
	\label{fig:injschema}
\end{figure}

\begin{figure}[!t]
	\centering
	\includegraphics[width=0.5\textwidth]{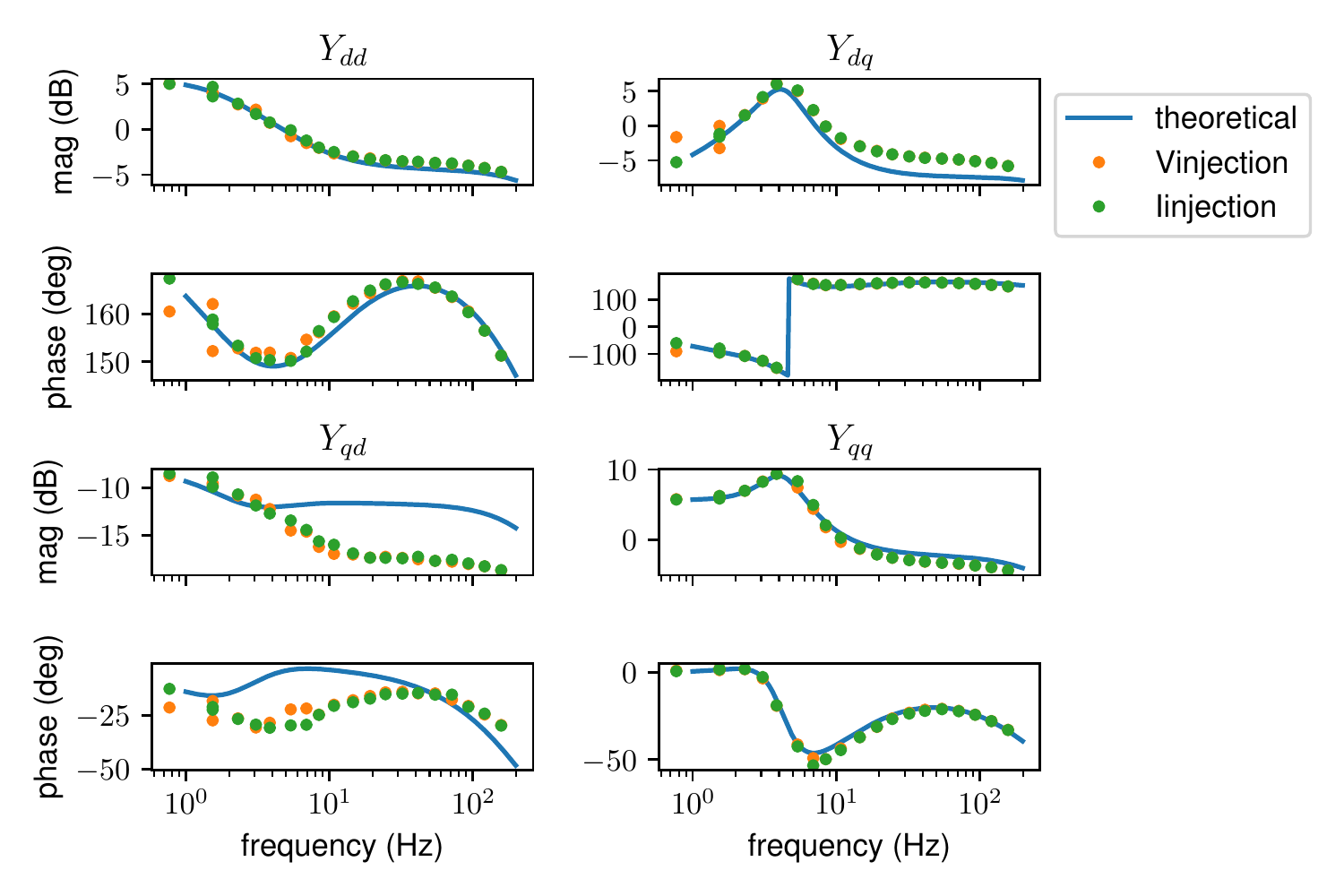}
	\caption{Comparison of the small signal based admittance and estimated admittance using voltage injection and current injection.}
	\label{fig:VinjIinjcompare}
\end{figure}

To measure the inverter admittance, a disturbance can be injected either to the inverter terminal voltage or current. The current and voltage disturbance injections require extra components to be added to the original circuit comprising the inverter and the grid, as shown in Fig. \ref{fig:injschema}.
The voltage disturbance element must be added in series, whereas the current disturbance element must be added in shunt.
Both voltage and current injections are used in the literature. Reference~\cite{ghe23} mentions that the voltage injection is less practical due to large currents which must pass through the voltage injection devices. However, voltage injection may still be practical especially if the component is being tested in a laboratory setup rather than in the field or for a software-based black-box model.
If testing the inverter using a controlled voltage source, the effect of the grid impedance (described later in Section \ref{sec:pllgrid}) can be minimized by using a source with a very small impedance. After removing the effect of the grid impedance, both voltage and current injection-based measurements in PSCAD give the same impedance for the modeled inverter, as shown in Fig. \ref{fig:VinjIinjcompare}.
\subsection{Choice of the injection signal}
The chosen range of frequencies of interest is 1-200 Hz since the modeled inverter uses an average switching model. A very simple signal to be injected is the sinusoid of a particular frequency. Reference~\cite{ghe21} provides a good comparison between some of the advanced injection signals proposed in the literature, and \cite{hgthesis} discusses this choice for the estimation of inverter admittance/impedance. The advanced injections mentioned in these papers aim to inject multiple frequencies at once so as to reduce the time required for the identification. Due to the restriction on magnitudes, multi-frequency inputs may not be very accurate unless designed properly. Further, \cite{ghe49} states that in some cases the single sinusoid injection leads to a more accurate estimation. Hence, in the present work, a disturbance containing a single frequency is applied at a time. The phase angle corresponding to the $dq$ frame is needed to perform the $dq$ to $abc$ conversion. Here, the same angle obtained from the PLL from the measurement device is used. It is shown in the literature that a difference in the small signal estimate of the phase angle with the actual phase angle does not lead to an inaccurate assessment of the admittance model~\cite{ghe41}.
\subsection{Magnitude and duration of the injection}
\begin{figure}[!t]
	\centering
	\includegraphics[width=0.5\textwidth]{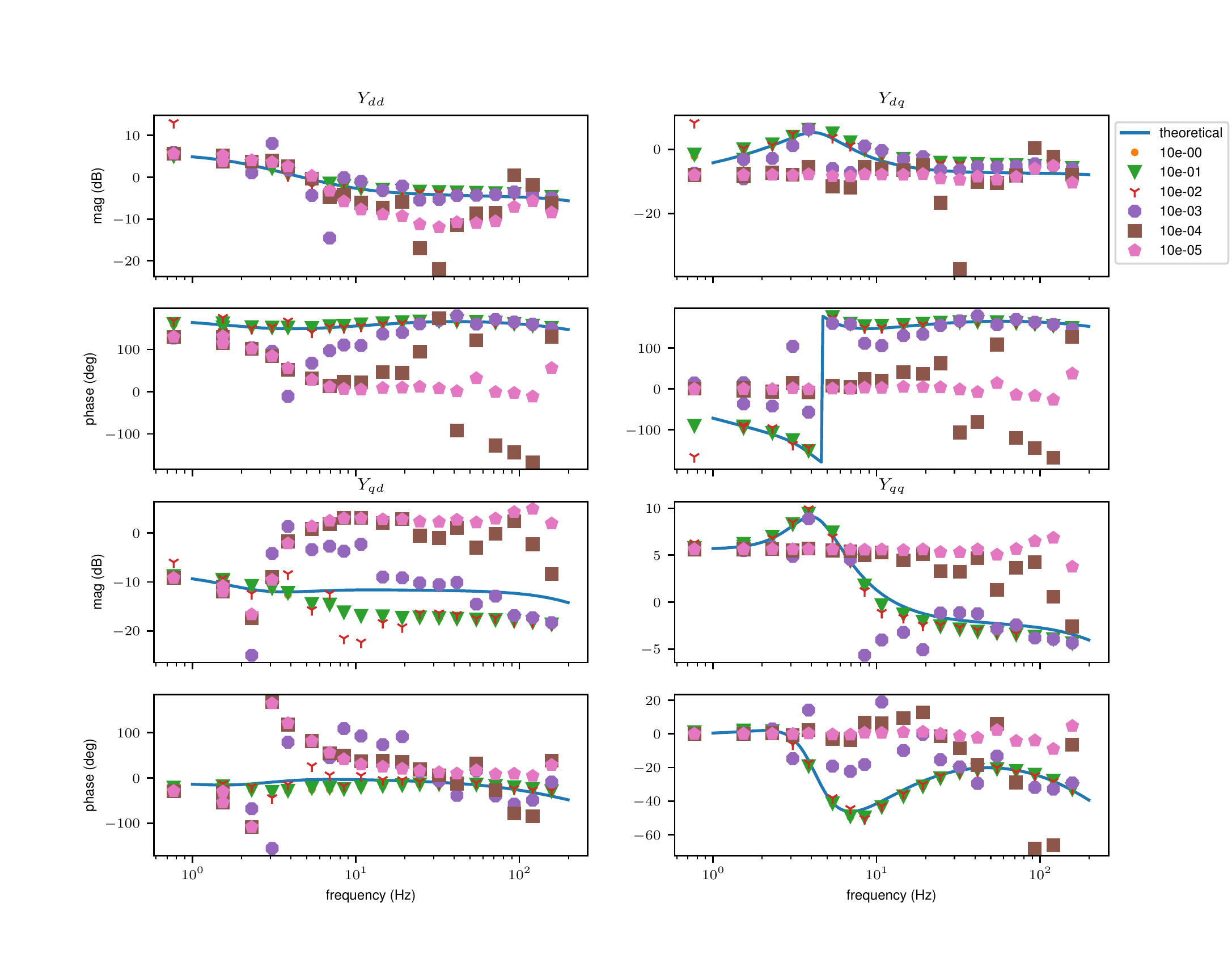}
	\caption{Impact of the injection magnitude on the estimated impedance for voltage injection. The magnitudes of injection shown are in kV.}
	\label{fig:VinjMagImpact}
\end{figure}

\begin{figure}[!t]
	\centering
	\includegraphics[width=0.5\textwidth]{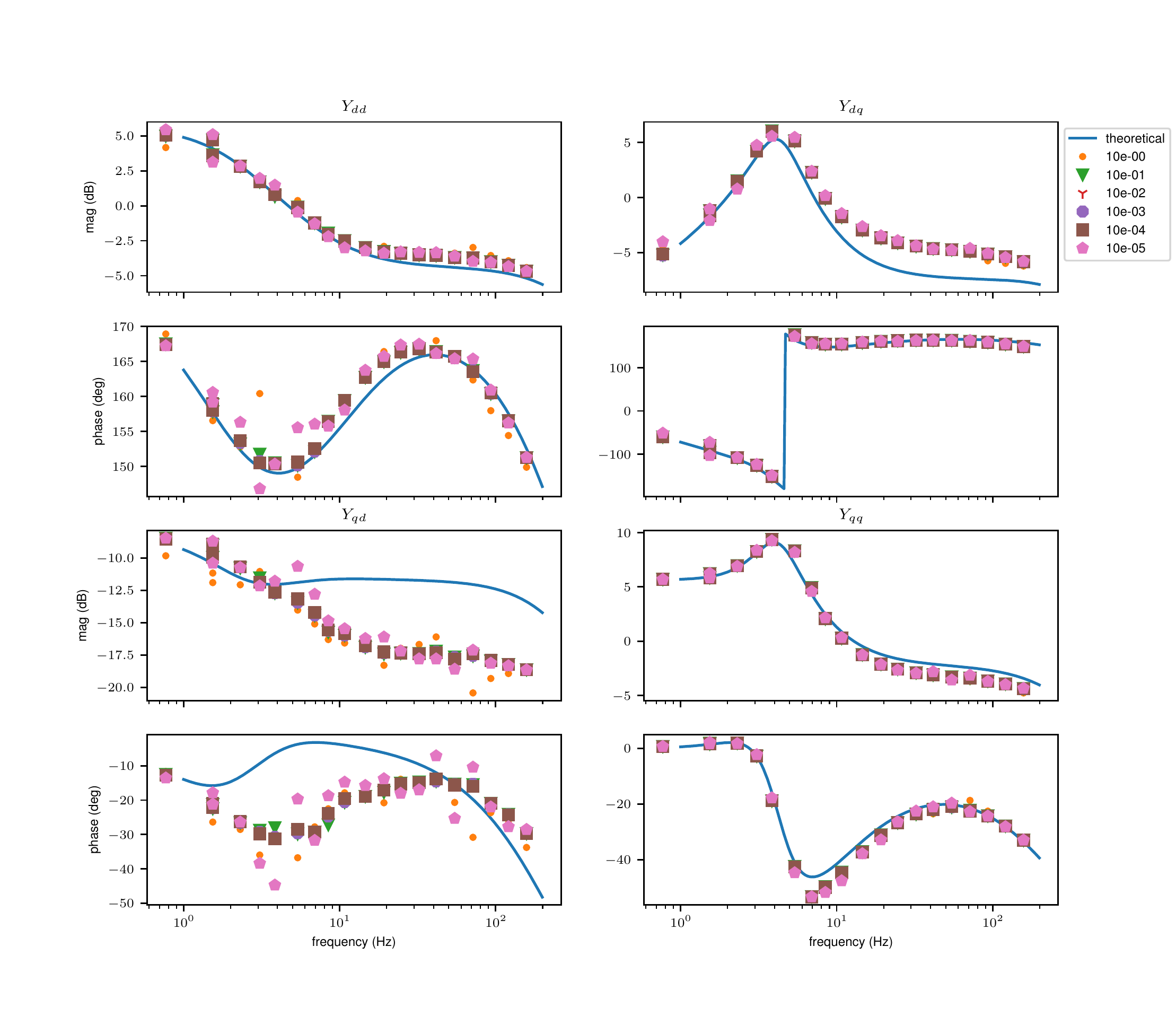}
	\caption{Impact of the injection magnitude on the estimated impedance for current injection.  The magnitudes of injection shown are in kA.}
	\label{fig:IinjMagImpact}
\end{figure}

If the injection magnitude is too small, the measured admittance may not be very accurate and may be prone to numerical errors and noise. However, too large of an injection will invalidate the assumption of linear response, and the measured admittance will not correspond to the steady-state operating point during the measurement. In some literature, the suggested magnitude is 5-10 percent of the nominal value~\cite{ghe41,ghe29,ghe30} based on experience. Some other sources use 0.1 percent~\cite{ghe28} or 0.025-0.002 p.u.~\cite{ghe49}. However, the injection magnitude is usually simply mentioned; hence, a study was conducted for the modeled inverter using both current and voltage injections with different injection magnitudes. It was found that too large or too small an injection leads to incorrect results, as shown in Fig. \ref{fig:VinjMagImpact} and Fig. \ref{fig:IinjMagImpact}. For the current injection, for example, a magnitude of $10^{-5}$ kA is too small and causes errors during the estimation, whereas an injection magnitude of $1$ kA is too large and causes the small signal assumption to not be true (not shown in the figure). As a comparison, the inverter nominal voltage and current are 120 kV and close to 1 kA.
Aside from the magnitude of the injection, the duration of the measurement also needs to be chosen. A fixed duration is sometimes used, such as 10 cycles used in~\cite{ghe29}. However,~\cite{ghe49} mentions choosing the time window based on the injected frequency - a similar approach is followed here. The duration should be chosen such that several cycles of the injected frequency are captured. In the present simulation, for each frequency, measurements were captured corresponding to at least 10 cycles of the injected frequency.\\

\begin{figure}[!t]
	\centering
	\subfloat[]{\includegraphics[width=0.25\textwidth]{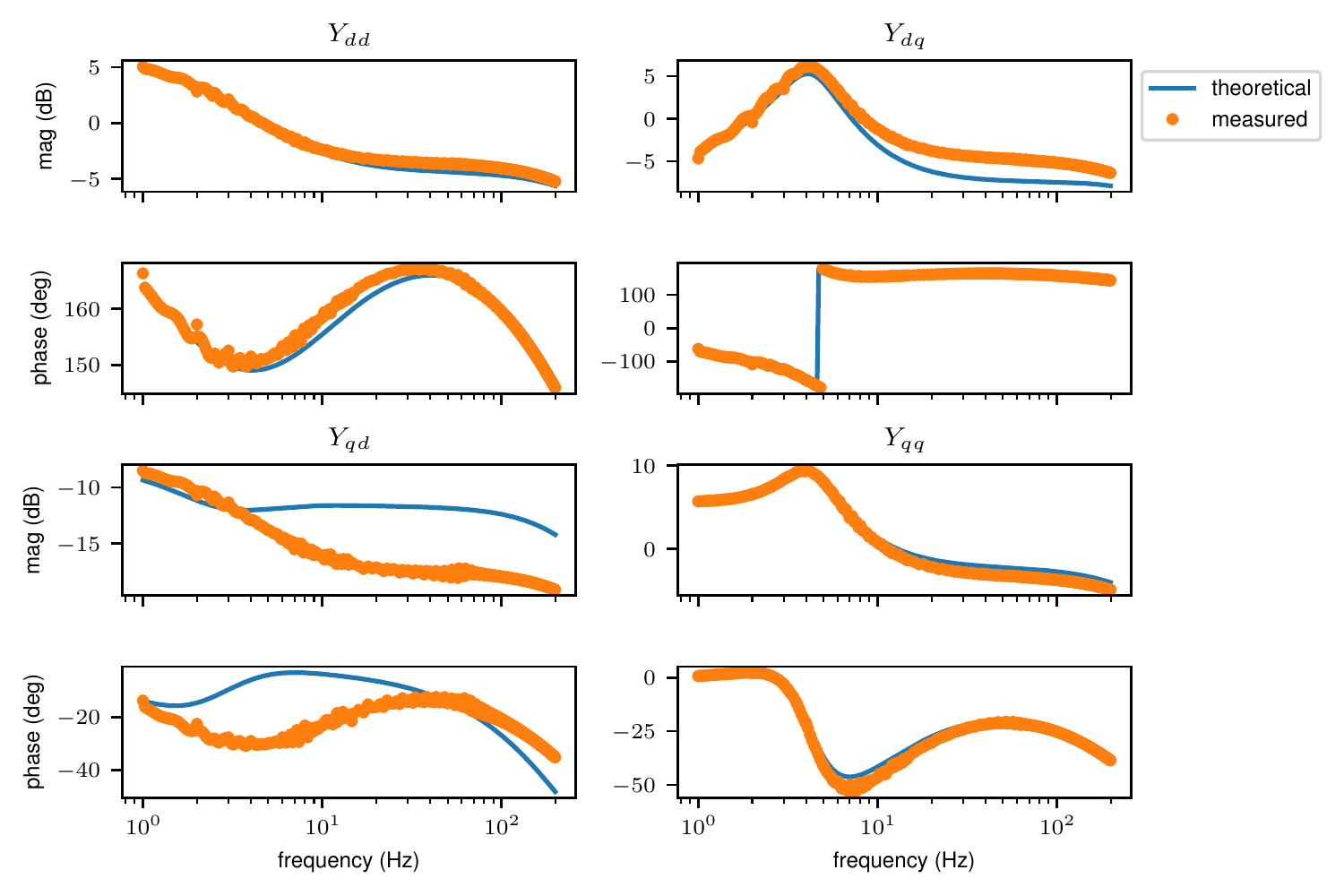}\label{fig:fftnoproperfreq}}
	\subfloat[]{\includegraphics[width=0.25\textwidth]{Figures2/Properfft.pdf}\label{fig:fftproperfreq}}
	\caption{The injection frequencies are not a part of the set of frequencies in FFT(a), leading to errors. When the injection frequencies are a part of the set of frequencies in FFT(b), the estimated admittance is smoother. The magnitude and phase plots at the top-left, bottom-left, top-right and bottom-right correspond to $Y_{dd}$, $Y_{qd}$, $Y_{dq}$ and $Y_{qq}$ respectively.}
\end{figure}

Another important aspect is to make sure that the measurement duration is such that the injected frequencies are part of the set of frequencies at which the Fast Fourier Transform (FFT) algorithm estimates the responses. If the injected frequencies are not a part of this set, the response may lead to additional errors during the calculations. Here Fig. \ref{fig:fftnoproperfreq} shows the comparison between the estimated and small signal modeling based transfer functions when the frequencies injected to the network are not a part of the frequencies used by FFT, whereas Fig. \ref{fig:fftproperfreq} shows the difference between the small signal based and estimated transfer functions when the injected frequencies are a part of the frequencies used by the FFT.\\

\subsection{Model fitting}
\begin{figure}[!t]
	\centering
	\subfloat[]{\includegraphics[width=0.4\textwidth]{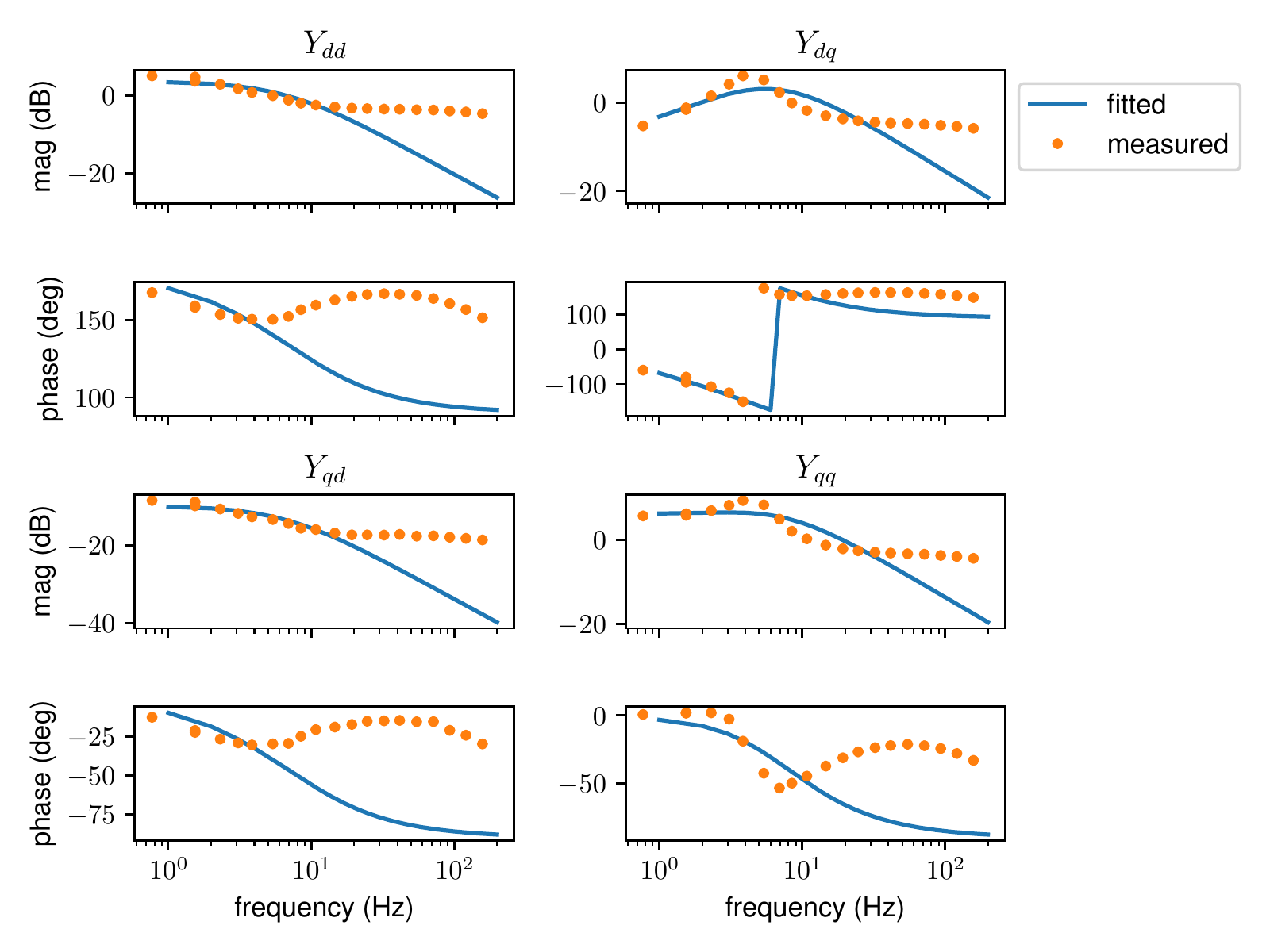}\label{fig:vfunderfit_plot}}
	
	\subfloat[]{\includegraphics[width=0.4\textwidth]{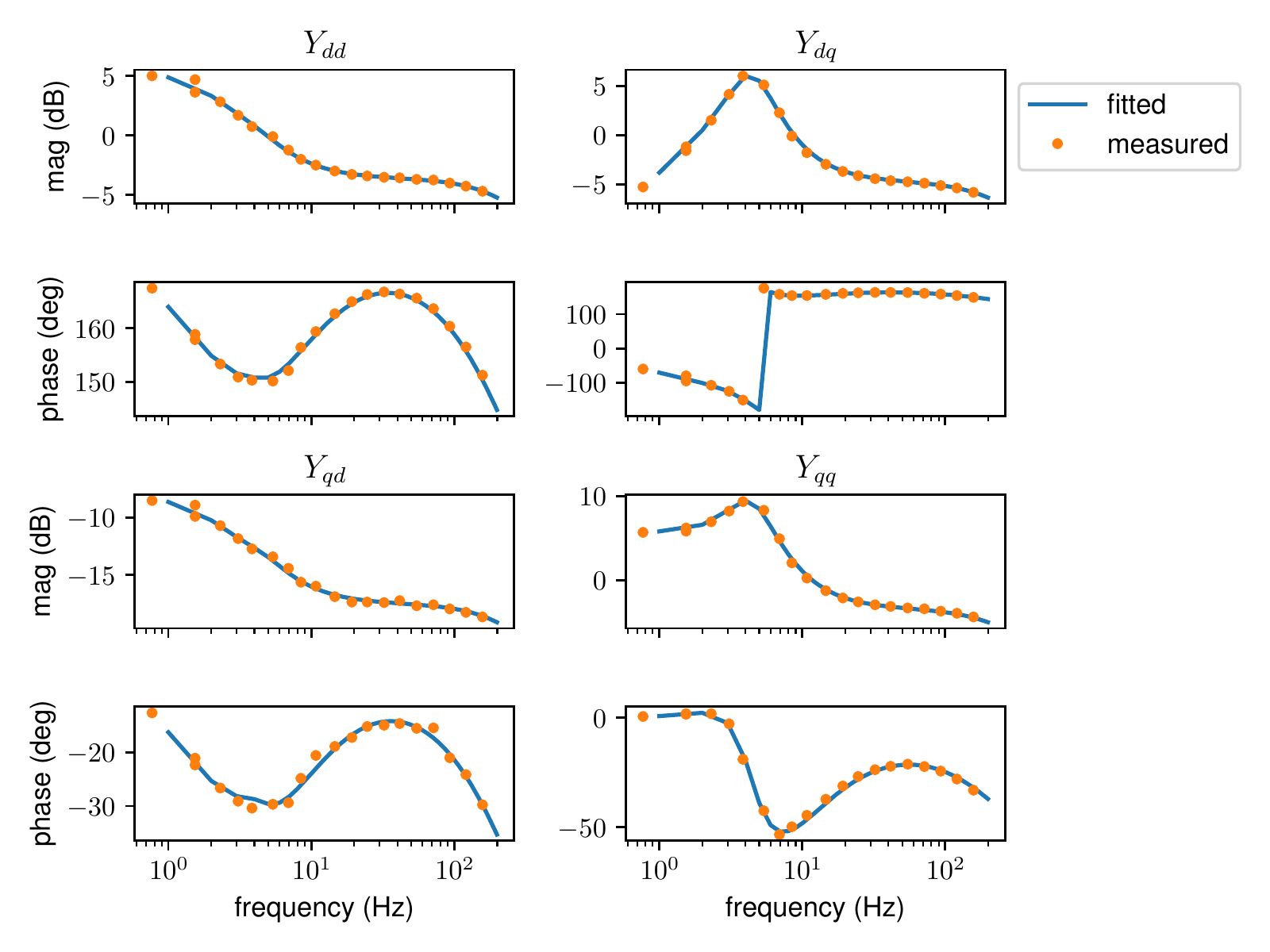}\label{fig:vfproperfit_plot}}
	
	\subfloat[]{\includegraphics[width=0.4\textwidth]{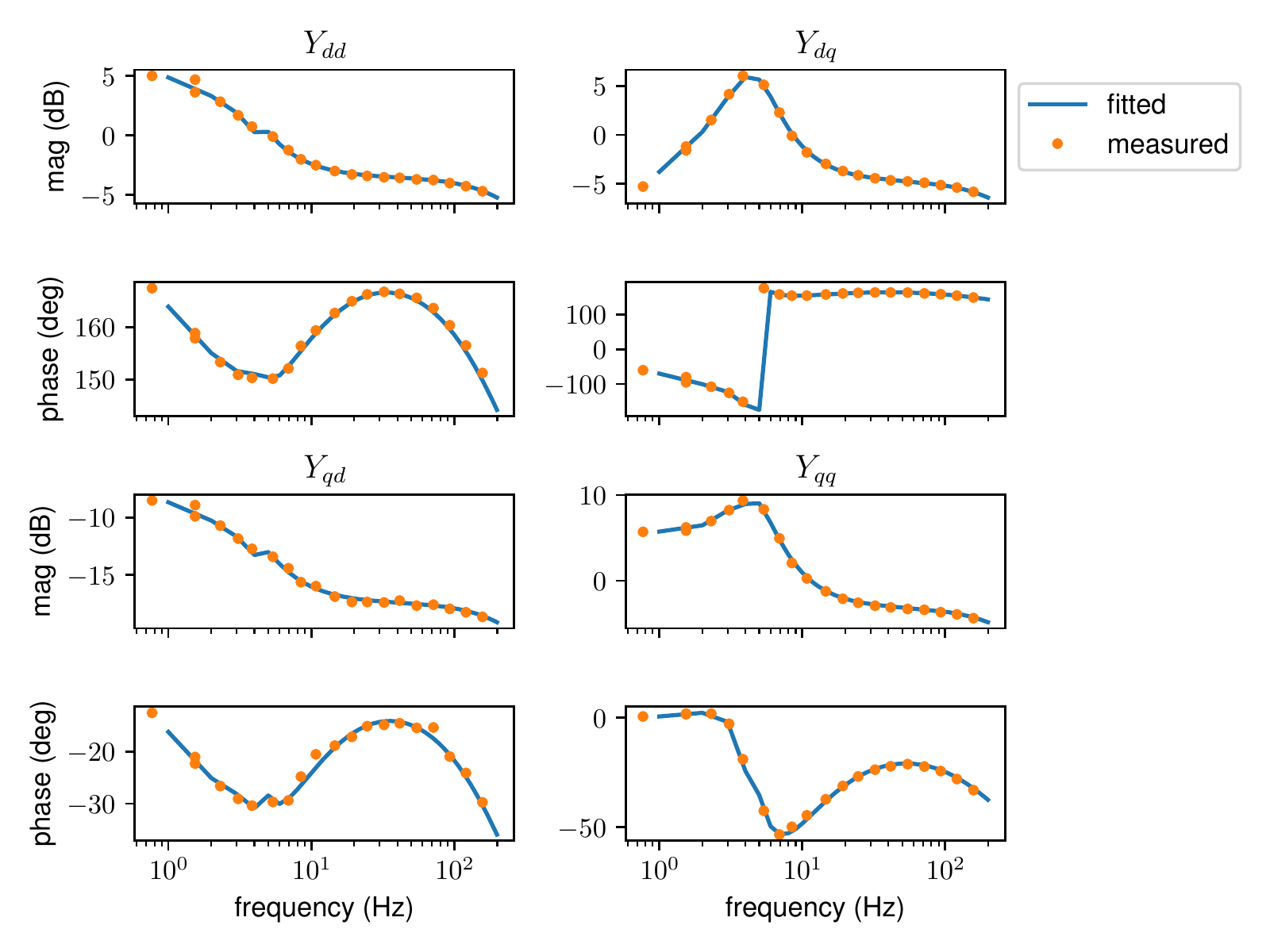}\label{fig:vfoverfit_plot}}
	\caption{Comparison of the measurement-based and fitted model frequency response, underfitted case (a), properly fitted case (b), and overfitted case (c).}
	
\end{figure}

Model fitting involves using the measurement-based admittance models at discrete frequencies to form/estimate a transfer function. The vector fitting approach from~\cite{si012} is used in this paper to analyze the impact of proper/improper fitting. This algorithm is available from a python package, `scikit-rf'. While performing vector fitting using the described algorithm, the user must specify the initial number of poles. Choosing the starting poles is an important decision while running the vector fit. There are some approaches in the literature that aim to estimate the number of poles required for a good fit~\cite{ghe42}, but usually, the approach suggested~\cite{ghe51} is to gradually increase the number of poles till the model reaches a good fit - so as to keep the number of fitted poles to the minimum. Fig. \ref{fig:vfproperfit_plot} shows a proper fit, whereas Fig. \ref{fig:vfunderfit_plot} shows a case of underfitting with lesser number of initial poles and Fig. \ref{fig:vfoverfit_plot} shows a case of overfitting. Note that to assess the sufficiency/accuracy of a fit, the modeler's judgment based on experience may be required. The modeler may also use other parameters such as the residue between the fitted and the measurement-based model for this judgment.\\

\section{Removing the influence of the PLL and the grid}\label{sec:pllgrid}
In addition to various aspects mentioned in the previous section, the influences of the measurement PLL and the grid impedance must be removed to create an accurate measurement-based model.
\subsection{Removing the influence of the measurement PLL}\label{subs:removePLL}
\begin{figure}[!t]
	\centering
	\includegraphics[width=0.5\textwidth]{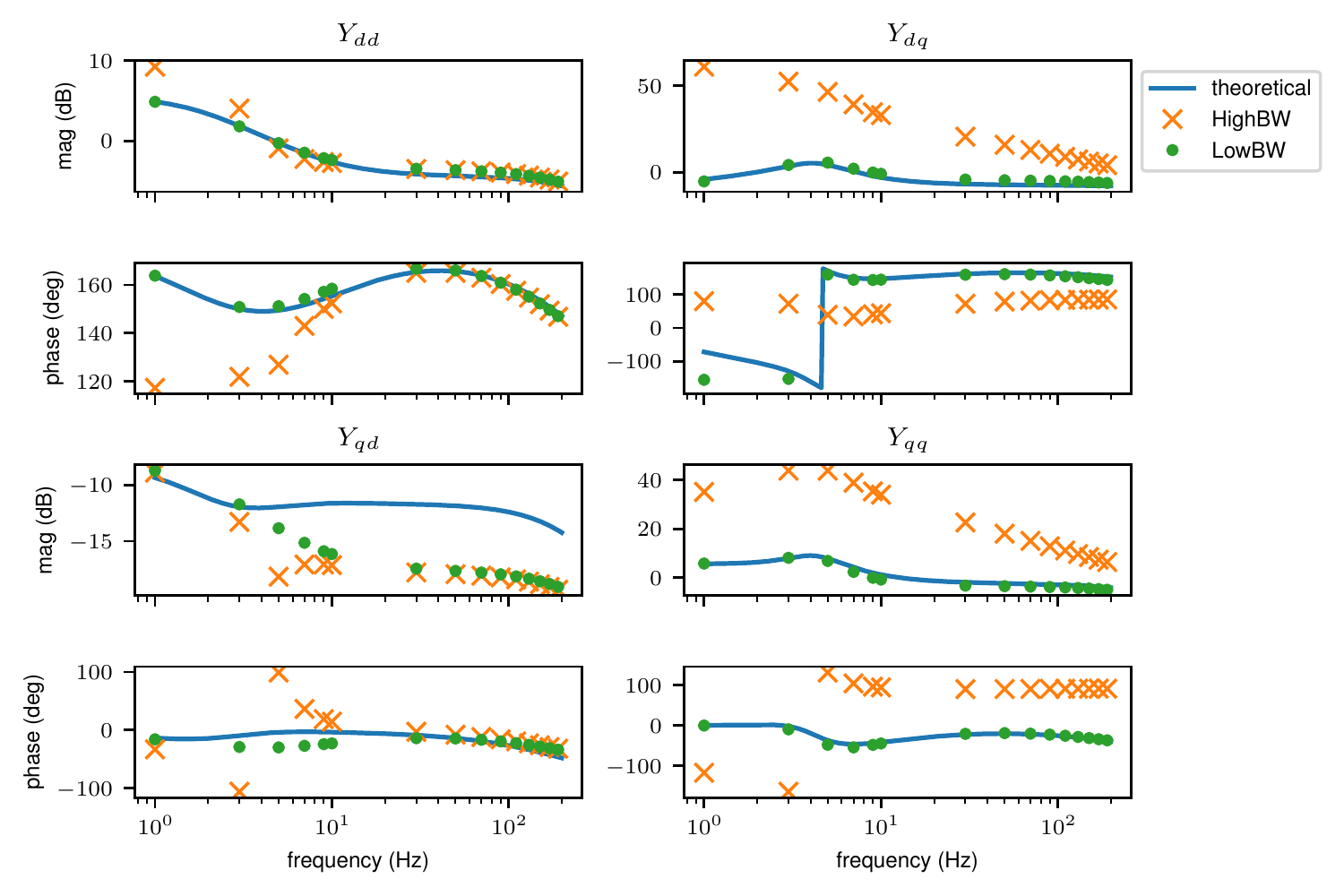}
	\caption{A high bandwidth measurement PLL results in an error in the estimated admittance, whereas using a low bandwidth measurement PLL removes this impact of the PLL.}
	\label{fig:lowBWPLL}
\end{figure}

\begin{figure}[!t]
	\centering
	\begin{circuitikz}[american voltages, american currents]
		\draw (-0.3,0) node[]{$V_q$};
		\draw[-latex] (0,0) to [twoport,t=PI,>] ++(2,0);
		\draw (2.5,0) node[adder] (adder2){};
		\draw[-latex] (2.5,1) node[label=above:{$2\pi60$}] {}to (adder2.n);
		\draw (adder2.e) to ++(1,0) node[label=below:{$\omega_{pll}$}]{} to [amp,t=1,>] ++(2,0) node[label=right:{$\theta$}]{};
		\draw (1.8,0) to ++(0,3) to [amp,t=1,>] ++(2,0);
		\draw (4.5,3) to [twoport,t=HPF,>] ++(1.5,0) node[label=right:{$d\theta_{filtered}$}]{};
		\draw (3.8,3) to (4,3);
		\draw (4.2,4) node[label=above:{0.0}]{} to (4.2,3.3);
		\filldraw (4.2,3.3) circle(1pt){};
		\filldraw (4,3) circle(1pt){};
		\filldraw (4.5,3) circle(1pt){};
		\draw (4.5,3) to (4.2,3.3);
		\draw (4.2,2.7) node[]{Switch};
	\end{circuitikz}
	\caption{Block diagram of the PLL for small signal analysis}
	\label{fig:pllcorrection2}
\end{figure}
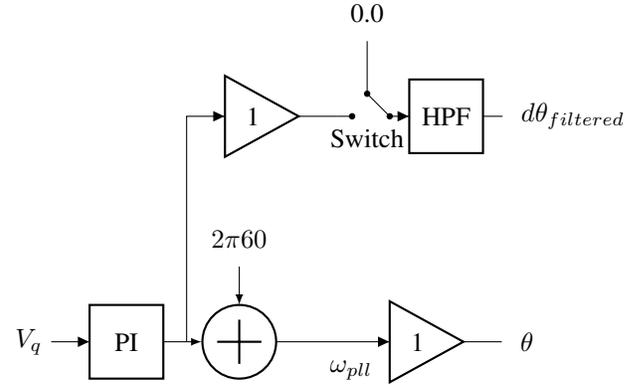

If a PLL is used to estimate the phase angle required by the $abc/dq$ conversion block for the measurement, it introduces an error in the estimation~\cite{ghe41}. This results in incorrect estimation of the inverter admittance particularly for the lower frequencies below PLL bandwidth, as shown in Fig. \ref{fig:lowBWPLL}. Originally the PLL bandwidth was around 45.77 Hz. One way to remove the influence of the PLL is to use a low bandwidth PLL~\cite{gfthesis}. Hence, a PLL of bandwidth 0.5 Hz was designed since the lowest injected frequency was 1 Hz. This removed the influence of PLL, as shown in Fig. \ref{fig:lowBWPLL}, however, a low bandwidth PLL results in increased time for the model to reach steady state and subsequently to adjust once the injection is applied.
Another approach to remove the influence of PLL is by recording the $d\theta_{filtered}$ variable during simulation~\cite{ghe41}. The correction to be applied in this case is described using the following equation:
\begin{align}
	\begin{pmatrix}
		X_{d}\\ X_{q}
	\end{pmatrix}
	=
	\begin{pmatrix}
		\cos(d\theta_{filtered}) & \sin(d\theta_{filtered})\\
		-\sin(d\theta_{filtered}) & \cos(d\theta_{filtered})
	\end{pmatrix}^{-1}
	\begin{pmatrix}
		X_{md} \\ X_{mq}
	\end{pmatrix},\label{eq:removepll}
\end{align}
where $d\theta_{filtered}$ is the PLL angle with respect to the grid $dq$ frame when a small signal disturbance is injected, $X_{mdq}$ are the measured quantities on the PLL $dq$ frame, and $X_{dq}$ are the original small signal quantities on the grid $dq$ frame. By applying this equation, the PLL can have a high bandwidth and hence is able to reach steady state faster. In this approach, to capture the variation in the tracked phase angle which is due to the injected disturbance,  the output of the PI controller in the measurement PLL is passed through an integrator and a high pass filter with a bandwidth less than the lowest injected frequency to obtain $d\theta_{filtered}$. However, even with this approach, because of the high pass filter, the recorded variable $d\theta$ takes a long time to reach steady state due to the variation at the start of the simulation. Hence, a further improvement (as compared to~\cite{ghe41}) is proposed in this paper by omitting the input to this filter and integrator at the start of the simulation till the transients while initializing the model have been mostly subsided using a switch, as shown in Fig. \ref{fig:pllcorrection2}.
Another approach can be to export the terminal voltage and current for the inverter in the $abc$ frame and convert it to the $dq$ frame without PLL - in this case, the reference angle for the $abc/dq$ conversion is computed by looking at the phase of the fundamental frequency by taking the FFT of the $abc$ voltage signals~\cite{ghe27}.
\subsection{Removing the influence of the grid}
\begin{figure}[!t]
	\centering
	\begin{circuitikz}[american voltages, american currents]
		\draw (0,0) node[tlground]{} to [R=$Z_{inv}$] ++(0,2) to [vsource,l=$v_{inj}$] ++(2,0) to [R=$Z_g$]++(0,-2) node[tlground]{};
		\draw (4,0) node[tlground]{} to [european resistor=$Y_{inv}$] ++(0,2) to ++(3,0) to [european resistor=$Y_g$]++(0,-2) node[tlground]{};
		\draw (5.5,0) node[tlground]{} to [isource, l={$i_{inj}$}] ++(0,2);
	\end{circuitikz}
	\caption{Small signal circuit representations for the voltage and current injection}
	\label{fig:injsmallsig}
\end{figure}
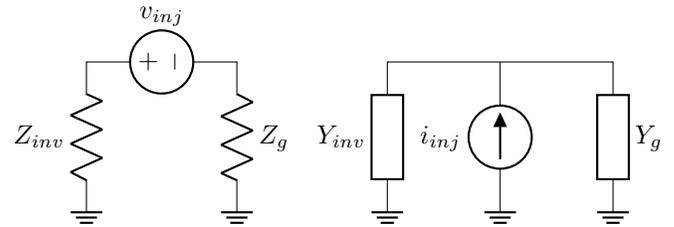

The inverter-grid circuits in small signal when applying voltage and current injections are shown in Fig. \ref{fig:injsmallsig}. It is seen that the injected small signal voltage/current disturbances will be divided across the inverter as well as grid impedance/admittance. Consequently, the measured impedance/admittance at the injection point will be a combination of the grid and inverter admittances/impedances. For voltage injection, by connecting a controlled source instead of the grid, the ``grid" impedance can be made very small, removing that influence - this is, however, not possible for current injection.\\

However, it is not necessary to remove the influence of the grid physically - it can also be removed by either accounting for it in the calculations~\cite{ghe41,ghe38} or simply by measuring the small signal voltage and current at the inverter terminal rather than at the injection location. In this case, a matrix equation must be formed using measurements from two sets of independent injections, and the inverter admittance at each frequency is estimated by solving $Y=I*U^{-1}$~\cite{gfthesis} rather than using \eqref{eq:Ydef}:
\begin{align}
	\begin{pmatrix}
		Y_{dd} & Y_{dq}\\
		Y_{qd} & Y_{qq}
	\end{pmatrix}
	=
	\begin{pmatrix}
		i_{md1} & i_{md2}\\
		i_{mq1} & i_{mq2}
	\end{pmatrix}
	\begin{pmatrix}
		v_{md1} & v_{md2}\\
		v_{mq1} & v_{mq2}
	\end{pmatrix}^{-1}\label{eq:Ycalc}
\end{align}
Here the subscript $m\{d,q\}\{1,2\}$ indicates the value measured on $d/q$ axis during the first/second injection.
\section{Small signal stability}\label{sec:stability}
\begin{figure}[!t]
	\centering
	\includegraphics[width=0.3\textwidth]{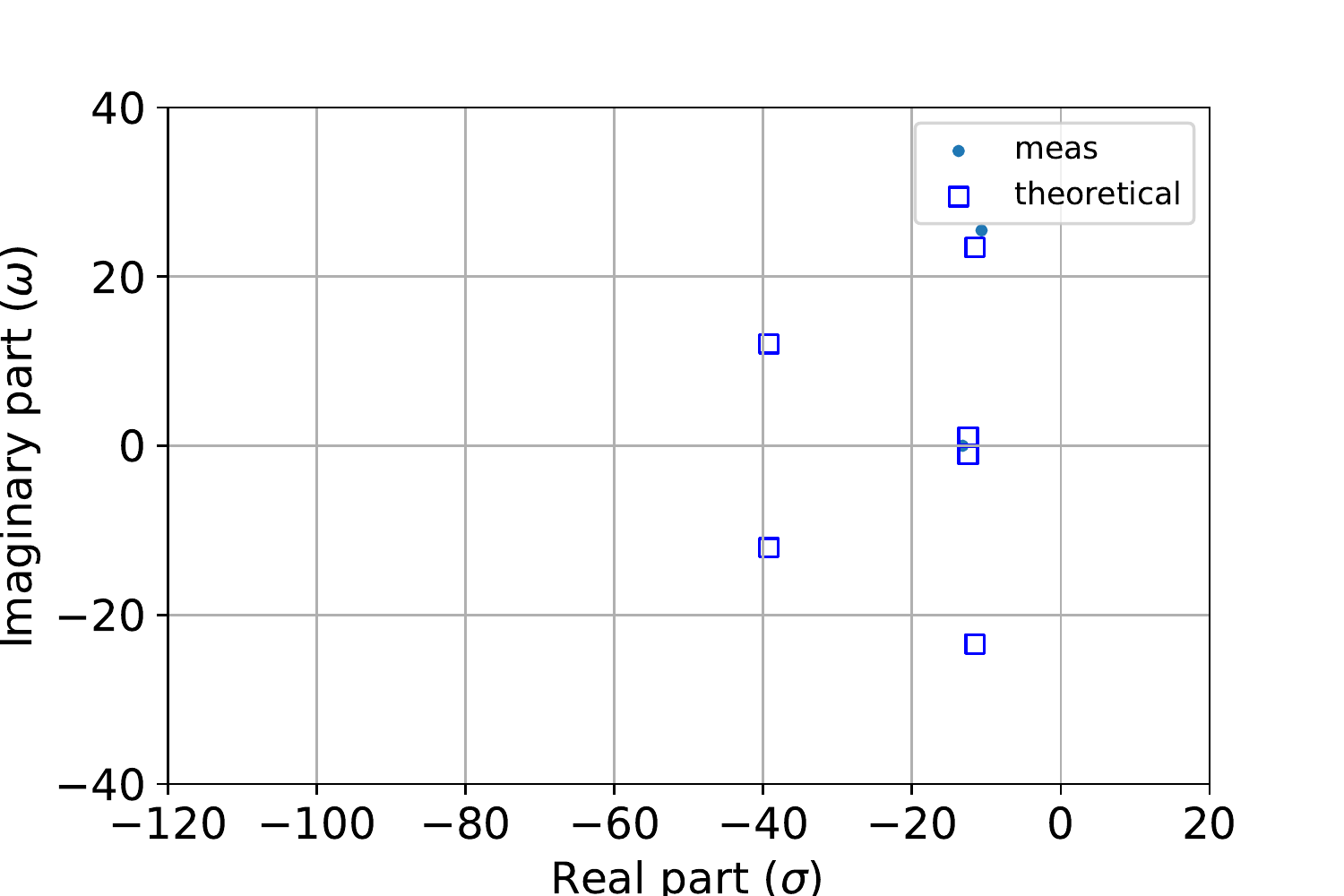}
	\caption{Comparison of the eigenvalues of the state matrix of the fitted model and the small signal based model.}
	\label{fig:vfproperfit_eig}
\end{figure}

One of the main purposes of estimating admittance as a transfer function is to be able to assess the stability of the inverter in question for various grid conditions. To compare between the small signal based and measurement-based models, the eigenvalues of the state matrix corresponding to the inverter from both approaches can be plotted together- if the fit is proper, the fitted poles will be close to some of the poles of the small signal model, as seen in Fig. \ref{fig:vfproperfit_eig}. Note that there are more poles in the small signal model which are not captured in the measurement-based model. However, the small signal stability characteristics of the measurement-based and small signal models for the considered IBR model are similar, as shown later.\\

It is known~\cite{usf} that inverter based resources can become unstable in weak grids. For the small signal based and measurement-based models, since these models are in the form of a transfer function, by combining them with the transfer function associated with the grid, it is possible to form a transfer function/state-space model of the combined inverter and grid system to assess the small signal stability for different values of grid impedance. However, for the PSCAD model, starting from a known stable grid impedance, after reaching steady state, the grid impedance is changed (with a corresponding change in the grid voltage so that the voltage at the inverter terminal is maintained at the same value) so as to push the system to a new impedance as well as to inject a small disturbance to the system to find whether the combined system of grid and inverter is stable at the new operating point. Note that it is important not to make too large of a change so that the operating point (the inverter terminal voltage and current) is preserved and small signal assumptions are valid.

\subsection{Forming the combined inverter-grid control system}\label{subsec:stabapproaches}
\begin{figure}[!t]
	\centering
	\begin{circuitikz}[american voltages, american currents]
		\draw[black] (0.0,0.0) rectangle(1.8,1);
		\draw (1.0,0.5) node[]{inverter};
		\draw (2.1,0.3) node{$v_t$};
		\draw (1.8,0.5) to [short, i=$i_t$] ++(2.0,0.0) 
		to[vsource, l=$v_{inj}$]++(1.0,0)
		to [R=$Z_g$] ++(2.0,0.0)
		to [vsourcesin, l=$v_g$] ++(2.0,0.0);
		\draw (8.8,0.5) node[ground]{};
	\end{circuitikz}
	\caption{Small signal circuit representation of the inverter connected to the grid with voltage injection based measurement}
	\label{fig:gridinvsmallsig}
\end{figure}
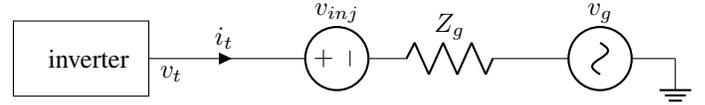

\begin{figure}[!t]
	\centering
	\begin{circuitikz}[american voltages, american currents]
		\draw[-latex] (0.0,0.0) node[label=left:{$v_{inj}$}]{} to ++(0.5,0);
		\draw (1.0,0) node[adder](add1){};
		\draw[-latex] (add1.e) to ++(0.5,0) node[label=below:{}]{} to ++(2,0) 
		to ++(1,0) node[label=right:{$v_t$}]{}
		++(-1,0) to ++(0,1) to [twoport,t=$Y_{inv}$,>] ++(-1.5,0) to [twoport,t=$Z_g$,>] ++(-1.5,0) to (add1.n);
	\end{circuitikz}
	\caption{Block diagram of the inverter plus grid system for small signal stability analysis}
	\label{fig:invgridsmallsignalanalysis2}
\end{figure}
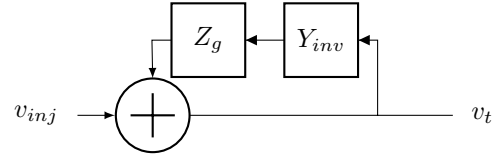

Different approaches are used in the literature to analyze the stability of an inverter model. Reference \cite{ss003} provides a good comparison between the Nyquist, Bode plot based and eigenvalue analysis. The grid and inverter network can be drawn as shown in Fig. \ref{fig:gridinvsmallsig}.
The control diagram involving the grid and the inverter can be drawn as shown in Fig. \ref{fig:invgridsmallsignalanalysis2}. Here $Y$ is the inverter admittance model obtained, and $Z_g$ is the grid impedance model given by
\begin{align}
	Z_{g}=\begin{pmatrix}R_g+sL_g & -w_0L_g\\w_0L_g & R_g+sL_g\end{pmatrix}.
\end{align}
The state-space model of the combined grid and inverter system is formulated using the python-control library for measurement-based and small signal based inverter models.
\subsection{Stability characteristics of the modeled inverter}\label{subsec:stabtest}
\begin{figure}[!t]
	\centering
	\subfloat[]{\includegraphics[width=0.25\textwidth]{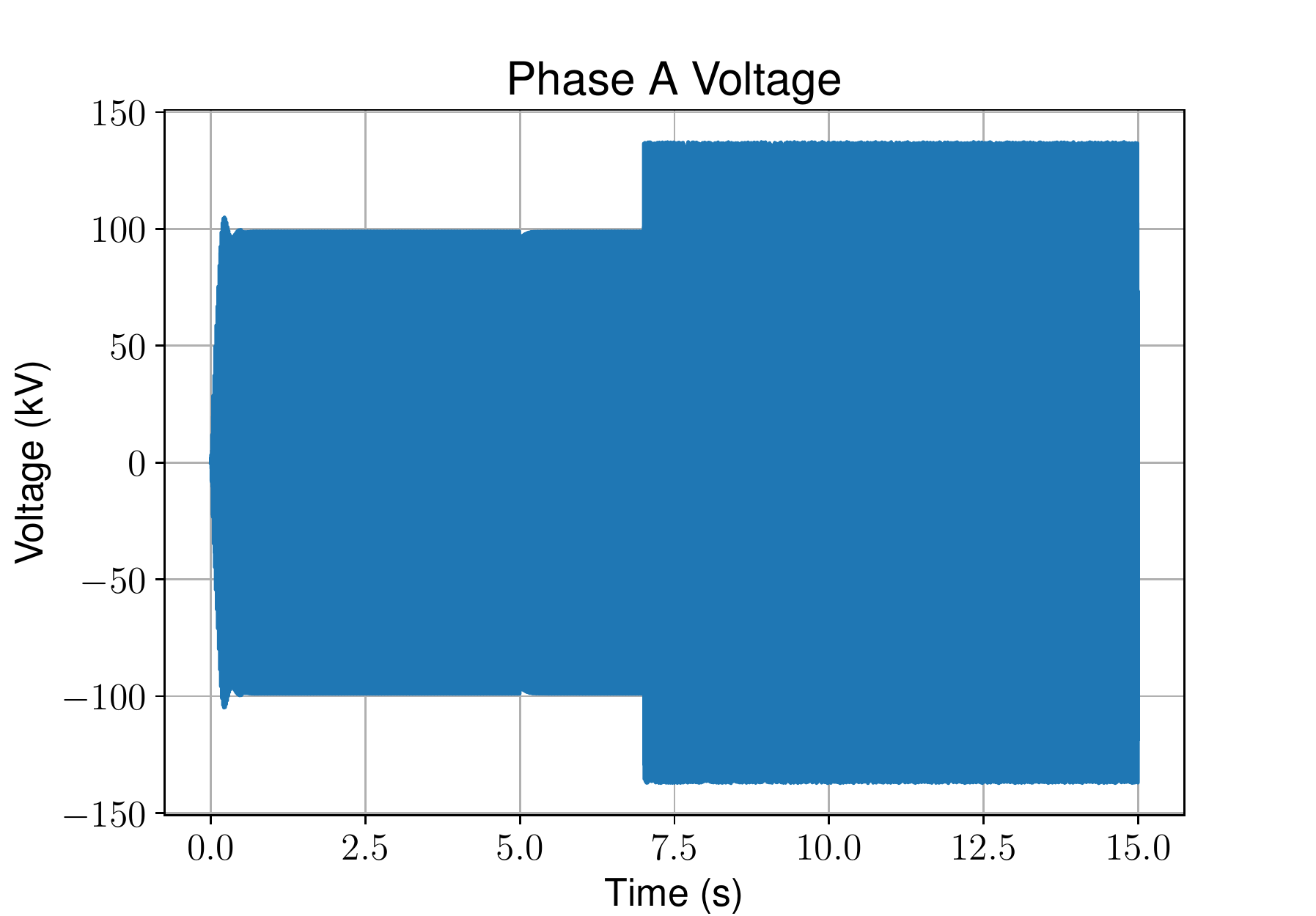}\label{fig:stabXV}}
	\subfloat[]{\includegraphics[width=0.25\textwidth]{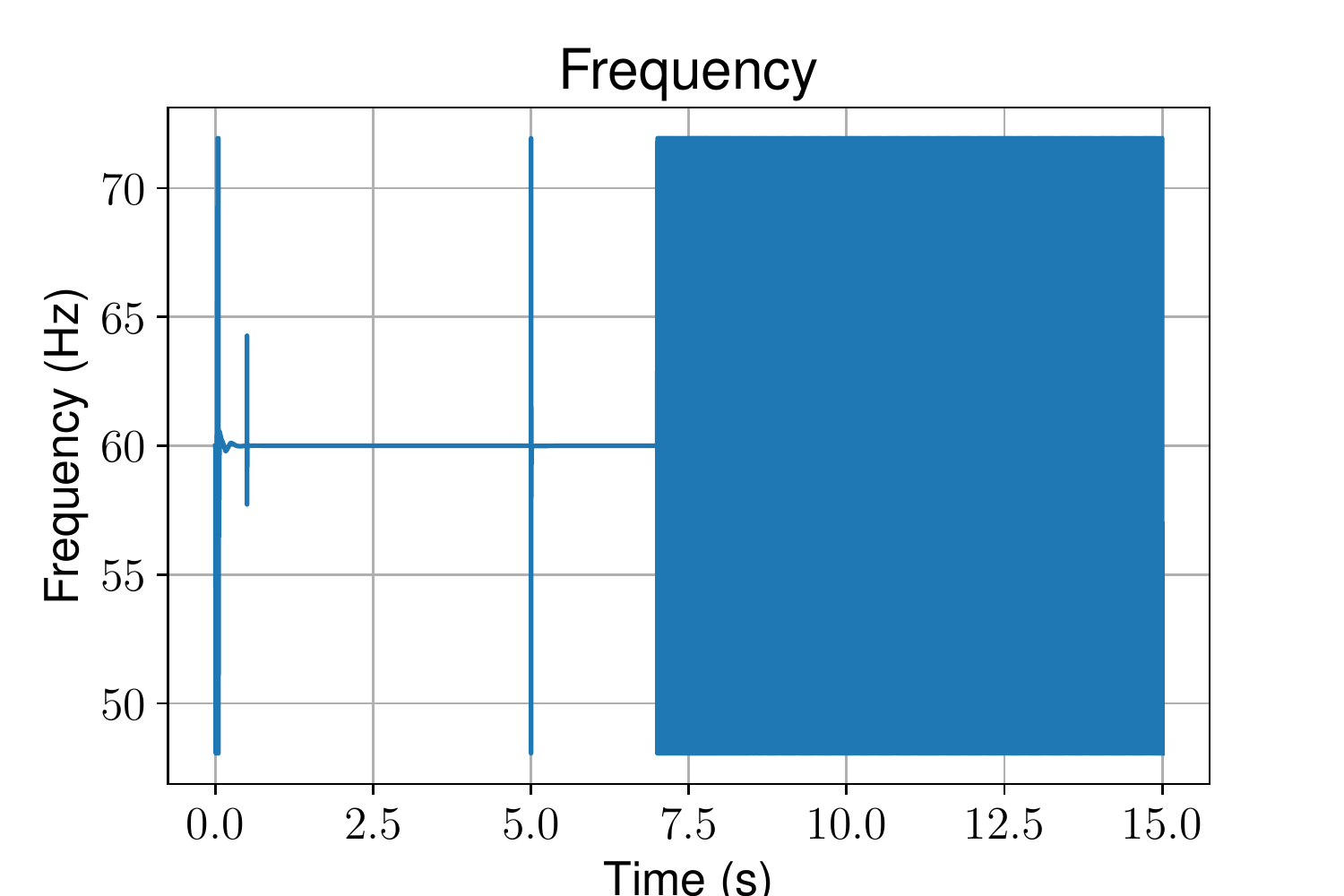}\label{fig:stabXf}}
	\caption{The voltage (a) and frequency (b) at the inverter terminal. The SCR of the  is changed at 5s from 4 to 1.6 and then to 1.5 at 7s.  The network $X/R$ ratio is kept at 6.}
\end{figure}

\begin{figure}[!t]
	\centering
	\subfloat[]{\includegraphics[width=0.25\textwidth]{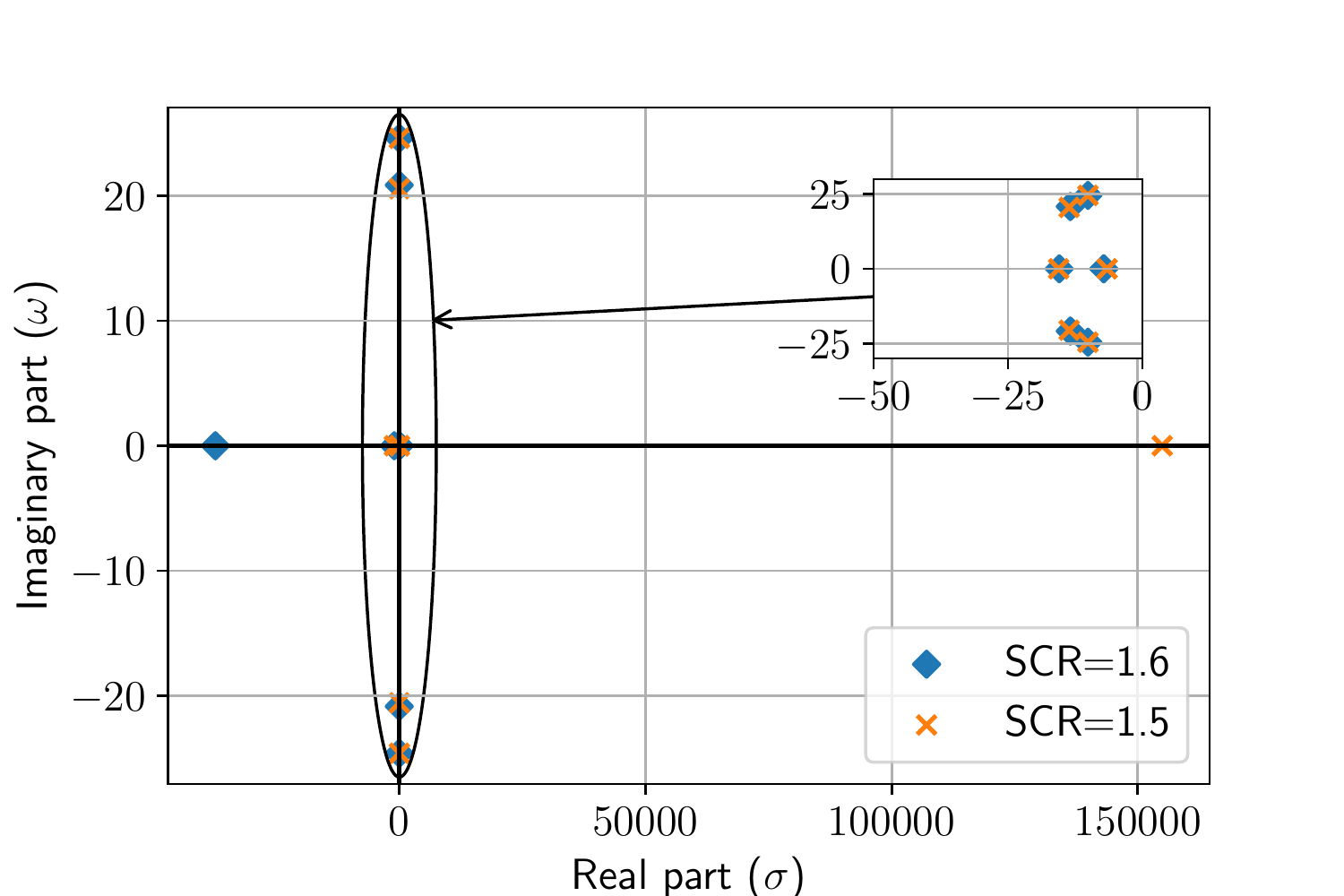}\label{fig:stab-meas}}
	\subfloat[]{\includegraphics[width=0.25\textwidth]{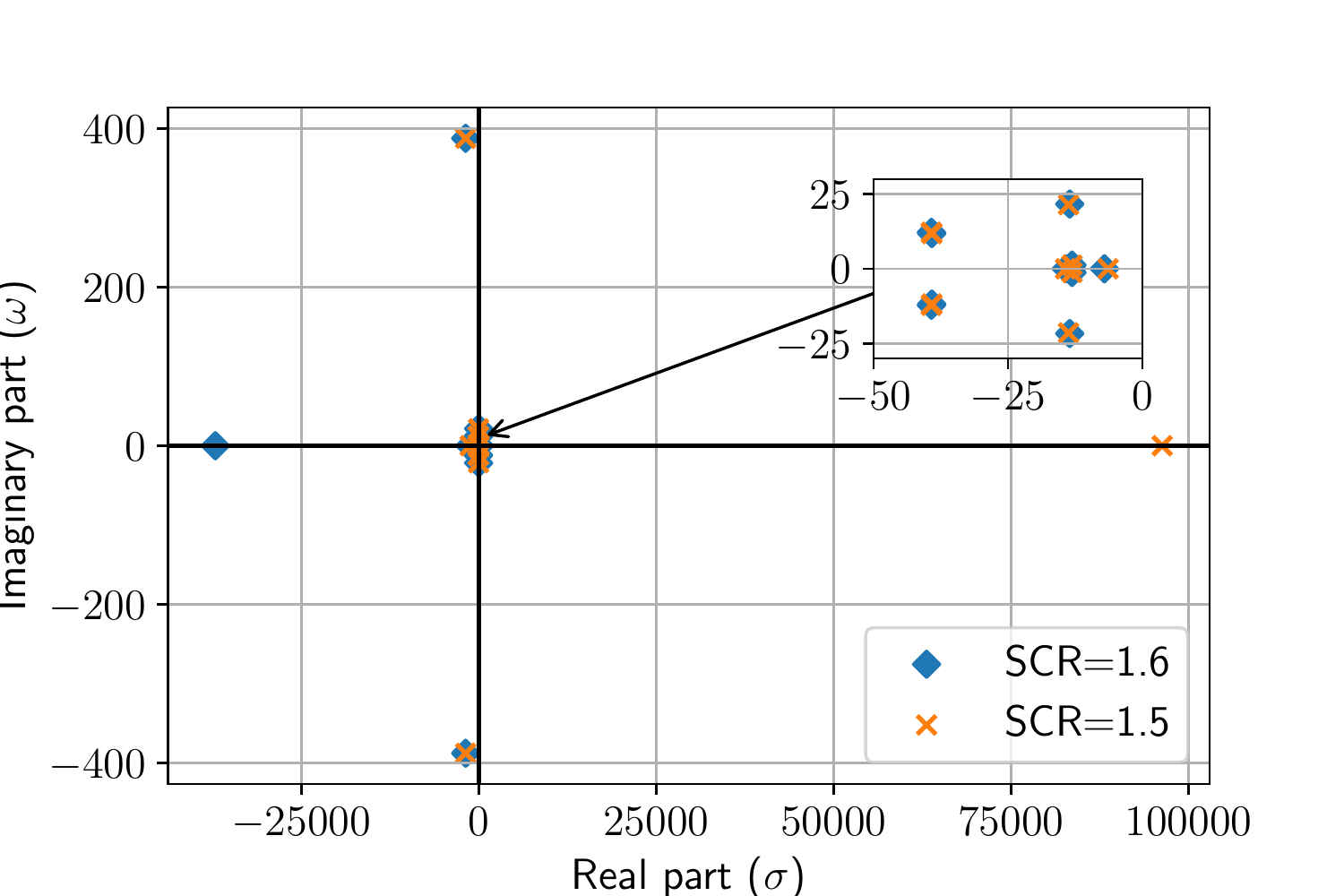}\label{fig:stab-ss}}
	\caption{The eigenvalues of the combined inverter-grid system for measurement-based (a) and small signal based (b) inverter models for the values of SCR of 1.5 and 1.6.}
\end{figure}

For the PSCAD model, it is found that for the $X/R$ value of 6.0, the PSCAD model becomes unstable at SCR=1.5. In this case, the initial operating point is obtained for SCR=1.6, and once the model reaches steady state, the SCR value is changed to 1.5 to discover that the model shows unstable behavior, as seen in Fig. \ref{fig:stabXV} and Fig. \ref{fig:stabXf}.
The eigenvalues of the combined inverter-grid system are plotted in Fig.~\ref{fig:stab-ss} and Fig.~\ref{fig:stab-meas} for small signal based and measurement-based models, respectively, for the values of SCR of 1.5 and 1.6, and $X/R=6.0$. Note that since the eigenvalues of the small signal and measurement-based inverter models are not identical (Fig.~\ref{fig:vfproperfit_eig}), the eigenvalues of the combined systems also show differences. However, both the small signal based and measurement-based inverter models combined with the grid impedance model show a stable behavior at SCR=1.6 and an unstable behavior at SCR=1.5, matching the stability characteristics of the PSCAD model.
\section{Conclusion}\label{sec:conclusion}
This paper explores creating a measurement-based admittance model for an inverter based resource. Such a model is created considering an inverter model as the black-box model as an illustration. Various factors affecting this process such as the selection of the variable and the signal for injection, the magnitude and duration of the injection, the model fitting process as well as the impact of the grid impedance and phase locked loop used during the measurement are examined, and it is found that while the effects of grid impedance and measurement phase locked loop can be removed by modification of the injection and model creation procedure, the other factors are also important and incorrect values/choices can lead to incorrectly estimated admittance. However, the values and choices regarding these factors appropriate for the illustrative model may not directly translate for all IBR black-box models. Despite the large number of challenges/choices, once the appropriate choices are identified for an IBR black-box model, the created measurement-based admittance model is able to adequately represent the small signal stability behavior of the black-box model for different grid impedances for the illustrative model. IBR admittance models could be useful for small signal stability studies involving large networks with many IBR and other models where the reduction in the computational resources required for conducting the stability studies is important. The measurement-based approach enables creating an admittance model for IBRs to be used in small signal stability studies while protecting the proprietary nature of the IBR controls. Hence, creating a measurement-based model is a process involving many decisions and more research might be needed to standardize this process.


\section*{Acknowledgment}

The authors would like to thank the team members at EPRI for providing valuable inputs and feedback. This work was funded by EPRI Research Program P173A: Modeling and Analytics with Emerging Technologies.



\bibliographystyle{IEEEtran}
\bibliography{PaperReferencesSushrut}
%



\end{document}